\documentclass[10pt,letterpaper]{article}
\usepackage[top=0.85in,left=2.75in,footskip=0.75in,marginparwidth=2in]{geometry}

\usepackage[utf8]{inputenc}

\usepackage{cite}

\usepackage{nameref,hyperref}

\usepackage[right]{lineno}

\usepackage{microtype}
\DisableLigatures[f]{encoding = *, family = * }

\raggedright
\usepackage{changepage}

\usepackage[aboveskip=1pt,labelfont=bf,labelsep=period,singlelinecheck=off]{caption}

\makeatletter
\renewcommand{\@biblabel}[1]{\quad#1.}
\makeatother

\usepackage{lastpage,fancyhdr,graphicx}
\usepackage{epstopdf}
\fancyheadoffset[L]{2.25in}
\fancyfootoffset[L]{2.25in}

\usepackage{color}

\definecolor{Gray}{gray}{.25}

\usepackage{graphicx}

\usepackage{sidecap}

\usepackage{wrapfig}
\usepackage[pscoord]{eso-pic}
\usepackage[fulladjust]{marginnote}
\reversemarginpar

\begin{document}
\vspace*{0.35in}

\begin{flushleft}
{\Large
\textbf\newline{Can the solar atmosphere generate very high energy cosmic rays?}
}
\newline
\\
Z. N. Osmanov \textsuperscript{1,2},
D. Kuridze \textsuperscript{3},
S.M. Mahajan \textsuperscript{4},
\\
\bigskip
$^1$ School of Physics, Free University of Tbilisi, 0183, Tbilisi, Georgia
\\
$^2$ E. Kharadze Georgian National Astrophysical Observatory, Abastumani 0301, Georgia
\\
$^3$ National Solar Observatory, 3665 Discovery Drive, Boulder, CO 80303, USA
\\
$^4$ Institute for Fusion Studies, The University of Texas at
Austin, Austin, TX 78712, USA

\end{flushleft}

\section*{Abstract}
The origin and acceleration of high-energy particles, constituting cosmic rays, is likely to remain an important topic in modern astrophysics. Among the two categories - galactic and solar cosmic rays - the latter are much less investigated. Primary source of solar cosmic ray particles are impulsive explosions of the magnetized plasma known as solar flares and coronal mass ejections. These particles, however,  are characterized by relatively low energies compared to their galactic counterparts. In this work, we explore resonance wave-wave (RWW) interaction between the polarized electromagnetic radiation emitted by the solar active regions, and the quantum waves associated with high-energy, relativistic electrons generated during solar flares. Mathematically, the RWW interaction problem boils down to analyzing a Klein-Gordon equation (spin less electrons) embedded in the electromagnetic field.  We find that RWW could accelerate the relativistic  electrons to enormous energies even comparable to  energies in the galactic cosmic rays.


\section{Introduction}

The solar atmosphere is a highly effective, astrophysical engine for particle acceleration.
It is well established that the primary manifestations of accelerated particles (such as electrons and protons)
in the solar atmosphere are flares and coronal mass ejections \cite{Benz2008}.
The flares, the most violent explosions in the solar system, release enormous amounts of energy ($\sim10^{32}$ ergs) impulsively through a  process, assumed to be driven by what is called magnetic reconnection. The energy released by the destruction of the free energy in the magnetic field accelerates particles (also known as non-thermal particles) that propagate along coronal magnetic field lines toward the solar surface and into the interplanetary medium \cite{Fletcher2011}.
The evidences of accelerated, relativistic/near-relativistic particles in the flaring solar corona are provided by remote-sensing observations, including $\gamma$-ray, hard X-ray/bremsstrahlung, and radio/gyrosynchrotron measurements \cite{Huds2019}.
The flare events are, often,  followed by dramatic enhancements in energetic particle fluxes, observed in in-situ measurements within the heliosphere, including at ground-level on Earth. 
These solar energetic particle (SEP) events, have also been referred to as solar cosmic rays.
The first detection of solar cosmic rays occurred through increased count rates in ionization chambers at ground level back in 1942 \cite{Forbush1946}.
The energy of solar cosmic ray particles are, typically, lower ( much lower) than galactic cosmic rays and range from a few MeV to several GeV  \cite{Huds2019}.

Several particle acceleration mechanisms, describing how free magnetic energy released during a flare can be converted into particle acceleration, have been discussed and studied over the decades. These mechanisms can be categorized into three main groups: (1) electric field acceleration, when particles are accelerated in strong electric fields generated during the flare; (2) 
first order Fermi-type (shock) accelerations \cite{fermi1,fermi2,catanese}; and 
(3) stochastic  (second-order Fermi) acceleration when particles gain energy through particle-wave interaction \cite[see chapter 11]{Aschwanden2004}.

In this paper, we will exploit a recently explored acceleration mechanism based on resonant wave-wave (RWW) interactions \cite{MA-16,MA-22}. This mechanism has already been applied  to study particle energization in the magnetospheres of radio pulsars and radio Active Galactic Nuclai (AGN) (\cite{resonAGN} -\cite{resonPulsar}); it was  shown that, under suitable conditions, RWW can boost particle energies to extremely high values - up to $10^{21}$ eV (AGN) and $10^{22}$ eV (pulsars).

The foundation work for the wave-wave resonance mechanism of acceleration was developed in \cite{MA-16,MA-22}, where a relativistic particle is described as a quantum wave - Klein Gordon (KG) or a Dirac wave. The interaction between an electron and a circularly polarized electromagnetic (EM) wave is investigated as an interaction between the KG and EM waves. {\it Thus the acceleration of a particle is equivalent to a KG wave gaining energy at the cost of radiation-an EM wave, generated, for instance, by a nearby astrophysical object}. The RWW works most efficiently when it operates on particles that are already fast.  It is worth noting that the same acceleration mechanism should be applicable to Sun-like stars and other flaring stars (such as low-mass M-dwarfs, as well as young flaring stars). However, detecting cosmic rays from other stars is much more challenging due to the low flux rate and the possible overlap with other cosmic ray sources.

In what follows, we will study the re-acceleration (via RWW) of already relativistic particles generated, for instance, in solar flares; the source of the electromagnetic fields needed for the second stage of acceleration also originate in the sun.  
The paper is arranged as follows: 1) In Section 2 , we summarize the relevant mathematical framework of RWW, 2) Section 3 describes the solar coronal environment, and parameters for efficient RWW, and 3) Sections (4 -5) present and summarize the main results of the study,


\section{Mathematical framework}
Since  the relativistic particles will be described via the quantum KG equation (see Appendix 1), we will begin with some relativistic preliminaries. The  group velocity of the KG wave, corresponding to a particle with mass $m$, 
\begin{equation}
\label{vg} 
\upsilon_g = \frac{\partial E}{\partial P} = \frac{P}{\left(P^2+m^2\right)^{1/2}},
\end{equation}
tends to the speed of light ($c = 1$) when $P>>m$. Here $E$ and $P$ represent the particle's energy and momentum respectively. For a strong interaction with the EM wave (traveling with the speed of light), the initial KG (quantum) wave  must be associated with highly relativistic particles. Thus even among the energetic constituents of the solar corona \cite{Chen2020}, it is only the fastest (high Lorentz factor) that would really resonate effectively with the wave.


Following \cite{MA-22} and considering the circularly polarized electromagnetic wave $A^0 = A^z = 0$, $A^x = A\cos\left(\omega t-kz\right)$ and $A^y = -A\sin\left(\omega t-kz\right)$ propagating along the $z$-direction, we will start our analysis with (Appendix A)
\begin{equation}
\label{KG} 
\left(\partial^2_t-\partial_z^2+2qAK_{\perp}\cos\left(\omega t-kz\right)\right)\Psi
+\left(K_{\perp}^2+m^2+q^2A^2\right)\Psi = 0,
\end{equation}
the KG equation embedded in the specified EM field. In the preceding equation, $\omega$($k$) represents the frequency (wave number) of the wave, $q$ is the particle's charge and $K_{\perp}$ represents the perpendicular momentum (a constant due to the fact that the perpendicular directions are ignorable).  We will seek solutions that are phase coherent with the EM, that is, $\Psi= \Psi(\xi)$ with $\xi = \omega t-kz$. Equation (\ref{KG}), then, reduces to the well known Mathew equation
\begin{equation}
\label{mathew} 
\left(\omega^2-k^2\right)\frac{d^2\psi}{d\xi^2}+\left(\mu+\nu\cos\xi\right) = 0,
\end{equation}
where $\mu = K_{\perp}^2+m^2+q^2A^2$ and $\nu = 2qAK_{\perp}$. It is worth noting that in a tenuous plasma (near vacuum), $\omega^2-k^2$ tends to zero making (\ref{mathew}) a singular differential equation; for a given $\mu$ and $\nu$, the derivatives of $\Psi$ must become large. The  corresponding energy/momentum of the KG wave, therefore, must become commensurately large- the physical system undergoes a resonant enhancement.

After taking the WKB approximation \cite{wkb} and dispersion relation, $\omega^2-k^2 = \frac{\omega_p^2}{\sqrt{1+q^2A^2/m^2}}$, into account, one can show that the rate of energy gain in physical units is given by \cite{MA-22,resonAGN}
\begin{equation}
\label{rate2} 
 \frac{dE}{dt} \simeq eAK_{\perp}c\frac{\sqrt{2}\hbar\omega^2}{\omega_p}\frac{\left(1+\frac{e^2A^2}{m^2c^4}\right)^{1/4}}{\left(1+\frac{e^2A^2}{m^2c^4}+\frac{\hbar^2K_{\perp}^2c^2}{m^2c^4}\right)^{1/2}}, 
 \end{equation}
where $\omega_p = \sqrt{4\pi nq^2/m}$ is the plasma frequency and $n$ represents the number density of particles.

The resonant enhancement, if unchecked, guarantees the maximum possible relativistic factor of the order of \cite{MA-16,MA-22}
\begin{equation}
\label{gama} 
\gamma\simeq \frac{\omega}{\omega_p}\left(1+\frac{e^2A^2}{m^2c^4}\right)^{1/4}\left(1+\frac{e^2A^2}{m^2c^4}+\frac{\hbar^2K_{\perp}^2c^2}{m^2c^4}\right)^{1/2}
\end{equation}
%



\begin{figure*}
\centering
\includegraphics[width=13.7 cm]{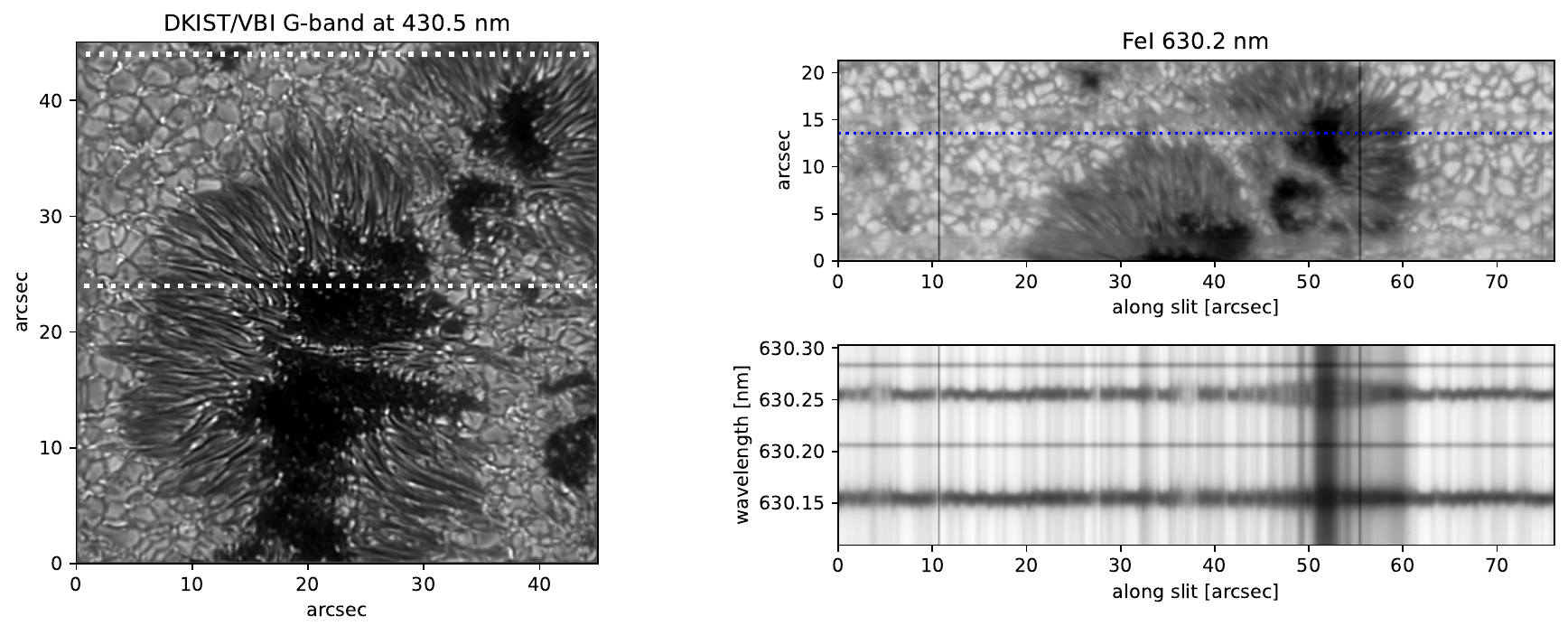}
\includegraphics[width=14. cm]{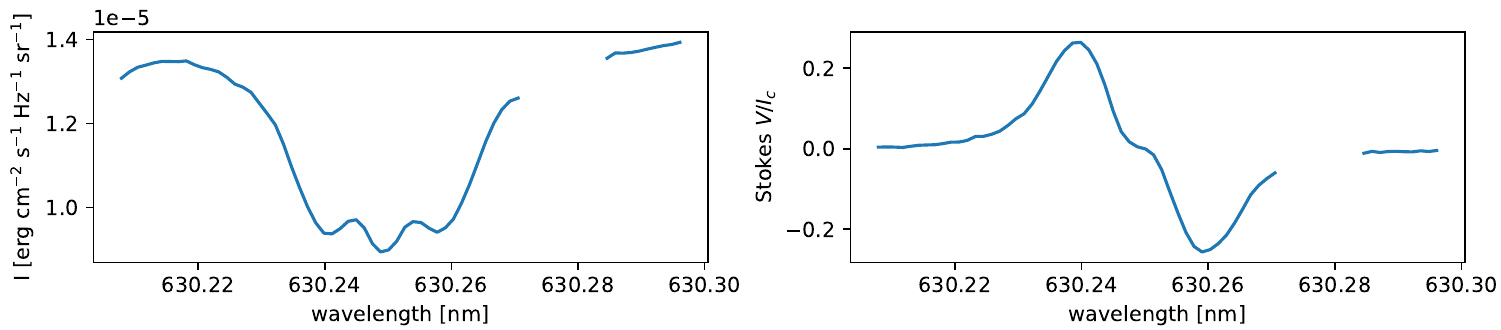}
\caption{Top left: Image of the sunspot in G-band observed with the Daniel K. Inouye Solar Telescope (DKIST)/Visible Broadband Imager (VBI) on October 16, 2023 in active region NOAA 13645.  White dotted lines mark area covered by ViSP spectral scan.  
Top right: DKIST/ViSP images of intensity (Stokes {\it{I}}) at 630.24~nm.
Middle right: Full spectra of Fe {\sc{I}} 630.1/630.2 nm lines along the slit position marked with blue dotted line in the top right panel. 
Bottom panels: A typical set of Fe Stokes {\it{I}} and {\it{V}} profiles of a
pixel located in a sunspot umbra at $52"$, $14"$ in the top right image.  
The O$_2$ telluric line has been removed from the observed spectra.}
\label{fig1}
\end{figure*}

\section{The Solar Active Region}

It is well known that the magnetic field of the solar active regions (ARs) induces polarized emission through the Zeeman effect in spectral lines formed across the entire electromagnetic spectrum from long radio wavelengths to GeV gamma rays.
Top panels of Figure~1 displays images of sunspot at G-band and Fe {\sc{I}} 630.2 nm observed with the Visible Broadband Imager \cite{Woger2021} and visible SpectroPolarimeter \cite{deWinj2014} on the Daniel K. Inouye Solar Telescope \cite{Rimmele2020} on 16 October 2023 in AR NOAA 13645.  
Middle right and bottom panels of Figure~1 shows Zeeman splitting of the Fe {\sc{I}} 630.1/630.2 nm  
spectra along the spectrograph slit and intensity (Stokes {\it{I}}) and circular polarization (Stokes {\it{V}}) profiles of the representative pixel located in the sunspot umbra. Specific intensity of the Fe {\sc{I}} 630.2 nm line in sunspot is around $I\approx 1.35\times 10^{-5}~\mathrm{erg~cm^{-2}~s^{-1}~Hz^{-1}~sr^{-1}}$ (Figure~1). Normalized Stokes {\it{V}} profile of this line indicates that around 30\% of the sunspot continuum emission is circularly polarized (Figure~1).

The active region of corona is a multi-thermal (from tens of thousands to tens of millions of Kelvin), tenuous and strongly magnetized plasma environment. 
The fundamental building blocks of coronal ARs are magnetic loops or open flux tubes, which are also the main contributors to coronal emissions.

Solar coronal polarimetry is challenging due to the low signal-to-noise ratios resulting from the weak magnetic field, hot and tenuous coronal environment and large Doppler width of spectral lines.
However, circular polarization of coronal emissions has been measured in various observations (see Table~\ref{tab}).
Reference \cite{Kuridze2019} reports measurements of  circular polarizations in infrared Ca $854.2$ nm line in coronal loops during the 10 September 2017 X8.3 class solar flare (second largest flare in solar cycle 24); magnetic field strength as high as 350 G (at heights up to 25 Mm above the solar limb) were inferred.

Coronal spectropolariemetry techniques, using the near-infrared coronal emission line of Fe XIII at 1074.7 nm \cite{Lin2000, Lin2004}, have found that the degree of circular polarization of Fe XIII at 1074.7 nm line is around $\sim$0.1\% of its disk-center intensity in AR corona. 
They inferred field strengths of 10 and 33 G in the two active regions at heights of 0.12, 0.15 $R_{\odot}$.

The circular polarization of low-frequency solar radio emissions has been recently explored by \cite{Morosan2022}. 
Their findings indicate that the average degree of circular polarization in the frequency range of 20/80 MHz varies between 10\% and 80\% during radio bursts.

Electron densities in the flare corona can range from  $\mathrm{10^{9}~to~10^{13}~cm^{-3}}$ \cite{Aschwanden2004}, depending on the strength of the flare and the temperature of the loops.
Analyses of microwave observations of the above mentioned X8.2 solar flare suggests that large flares can accelerate nearly all electrons in a large coronal volume to mildly relativistic and relativistic speeds \cite{Fleishman2022}. 
The estimated number density of electrons
with energies above 20 keV in \cite{Fleishman2022} is up to $\sim\mathrm{10^{10}~cm^{-3}}$.
Furthermore, it has been derived that the number density of the relativistic electrons with energy above 300 keV (up to $\sim 1$ MeV) during this event range from $10^{2} - 10^{4.7}$ cm$^{-3}$  \cite{Chen2020}.  \\

Although we have emphasized the circularly polarized radiation in the preceding
discussion, RWW can take place for arbitrary polarization; hence, simple analytical expressions
for energy gain (for instance) are possible for the circularly polarized waves and
comparison with observation becomes easier.


\section{Results}

The relativistic particles in the flaring corona are illuminated by polarized emissions originating from both the corona and the photosphere, creating a promising avenue for RWW interaction. Table 1 summarizes the parameters of radiation intensity across four different wavelength ranges mentioned in Section 3, which have been successfully used for coronal magnetic field diagnostics in several studies.


As an example, we consider flare corona at $H\sim 25$ Mm above the photosphere. Then by taking an average diameter, $D$, of the sunspot $\sim 20$ Mm into account (Figure~\ref{fig1}), one can estimate (order of magnitude) the corresponding solid angle $\Omega\simeq 2\pi\left(1-\frac{H^2}{\sqrt{H^2+D^2}}\right)\simeq 0.45$ sr;  the  flux $F_0\simeq 611.2$ erg cm$^{-2}$ s$^{-1}$ is, then, estimated after multiplying by the fraction of the polarized part $0.01$ for $\lambda_0 = 854.2$ nm (see Table 1).


For such a scenario, one can obtain, 
\begin{equation}
\label{gama1} 
\gamma\simeq 3.9\times 10^9\times \frac{\omega}{\omega_0}\times \left(\frac{10^{2} cm^{-3}}{n}\right)^{1/2},
\end{equation}
where $\omega_0 = 2\pi c/\lambda_0$, and the flux and the cyclic frequency are normalized by the values corresponding to Fe I $630.2$ nm line. This is the maximum value of the RWW boosted Lorentz factor. However if there existed a mechanism that saturates the acceleration process at a lower Lorentz factor, the latter will control the final energy.

In general, there are several limiting mechanisms, which might potentially impose strict constraints on the maximum achievable energies. The solar corona magnetic field, $B$, can vary from tens up to hundreds of Gauss and the charged relativistic particles will inevitably experience synchrotron losses with the rate \cite{rybicki}
\begin{equation}
\label{syn} 
P_{syn}\simeq\frac{2e^4B^2\gamma^2}{3m^2c^3}.
\end{equation}
As particles gain more and more energy (reaching higher $\gamma$) the synchrotron losses become stronger limiting the maximum allowed energy. The synchrotron limited $\gamma_{syn}^{max}$, obtained by balancing $dE/dt\simeq P_{syn}$, is estimated as
$$\gamma_{syn}^{max}\simeq\left(\frac{dE}{dt}\times \frac{3m^2c^3}{2e^4B^2}\right)^{1/2}\simeq 7\times 10^{8}\times\frac{350\; G}{B}\times$$
\begin{equation}
\label{gsyn} 
\times\left(\frac{\omega}{\omega_0}\right)^{1/2}\times\left(\frac{F}{F_0}\times\frac{10^{2} cm^{-3}}{n}\right)^{1/4},
\end{equation}
where we have taken into account the expression, $A = c\sqrt{cF}/\omega$, \cite{resonAGN,resonPulsar} 

Another mechanism that can potentially limit the energization process is the inverse Compton (IC) scattering when energetic electrons encounter soft thermal photons in the solar ambient. It is worth noting that the scattering will be more efficient with the radiation coming from the photosphere, than with the photons corresponding to the coronal region, because the number density of the latter is much smaller. In the Thomson regime, the IC cooling rate is given by \cite{rybicki}
\begin{equation}
\label{IC} 
P_{_{IC}}\simeq\sigma_Tc\gamma^2U,
\end{equation}
where $\sigma_T$ is the Thomson cross section, $U\simeq \sigma T^4R_{\odot}^2/\left( (R_{\odot}+H)^2c\right)$ is radiation energy density in the corona at the altitude $H$, $\sigma$ denotes the Stefan-Boltzmann constant, $T\simeq 5777$ K is the photospheric temperature and we assume that the radiation is black body. The balance condition yields
$$\gamma_{_{IC}}^{max}\simeq\left(\frac{R_{\odot}+H}{R_{\odot}T}\right)^2\times\left(\frac{dE/dt}{\sigma\sigma_T}\right)^{1/2}\simeq$$
\begin{equation}
\label{gIC} 
\simeq 4.9\times 10^{10}\times\left(\frac{\omega}{\omega_0}\right)^{1/2}\times\left(\frac{F}{F_0}\times\frac{10^{2} cm^{-3}}{n}\right)^{1/4}.
\end{equation}

\begin{table*}
    \centering
    \caption{Parameters of intensity for selected wavelengths/frequencies used for polarization measurements in the Sun.}
    \label{tab}
    \begin{tabular}{l lc lcc lccc}
    \hline 
    Wavelength/ & Intensity  & Circular~polarization  & Coronal magnetic field \\ 
    frequencies  &  $\mathrm{[erg~cm^{-2}~s^{-1}~sr^{-1}]}$ & (Stokes V/I$_{c} ~[\%]$)  &  strength [Gauss] \\ 
    \hline
    Fe {\sc{i}} 6302 {\AA}$^{a}$                  & $1.02\times10^{6}$ &  25 &  N/A \\
    Ca {\sc{ii}} 8542 {\AA}$^{b,c}$               & $1.36\times10^{5}$ &  0.5 - 2  & 350 \\
    Fe {\sc{xiii}} 10747 {\AA}$^{d,e}$                  & $1.35\times10^{2}$   &  0.1 & 10 - 30 \\
    20 – 80 MHz$^{f}$                  & $2.40\times10^{-12}$ $^{g}$  &  20 - 80 & N/A \\
    \hline
    \end{tabular}
    \vspace{1mm}
    
$^{a}$Figure~\ref{fig2}, $^{b}$\cite{Kuridze2019}, $^{c}$\cite{Koza2019}, $^{d}$\cite{Lin2000}, $^{e}$Chianti atomic line database \cite{Landi2013}, $^{f}$\cite{Morosan2022}, $^{g}$The unit of the radio flux is $\mathrm{[erg~cm^{-2}~s^{-1}]}$.
\end{table*}

One should emphasise that the IC scattering in the Thomson regime works only when $\gamma_{_{IC}}^{max}\epsilon_{ph}/(mc^2)<<1$, where $\epsilon_{ph}\sim k_BT$ is the photon energy. It is straightforward to check that, for all lines shown on Table 1, the above mentioned condition is violated.The scattering, therefore, occurs in the so-called Klein-Nishina regime. The corresponding expression for cooling power \cite{blum}
\begin{equation}
\label{KN} 
P_{_{KN}}\simeq\frac{\sigma\left(mckT\right)^2}{16\hbar^3}\left(\ln\frac{4\gamma kT}{mc^2}-1.981\right),
\end{equation} 
will lead to the maximum Lorentz factor 
\begin{equation}
\label{gKN} 
\gamma_{_{KN}}^{max}\simeq\frac{mc^2}{4k_BT}\exp\left\{\frac{16\hbar^3}{\sigma\left(mck_BT\right)^2}+1.981\right\},
\end{equation} 
where $\hbar$ denotes the Planck's constant.The preceding value of   $\gamma_{_{KN}}^{max}$ is unrealistically high. The IC scattering, therefore, does not impose any constraints on resonant energization.

\begin{figure}
\resizebox{\hsize}{!}{\includegraphics[angle=0]{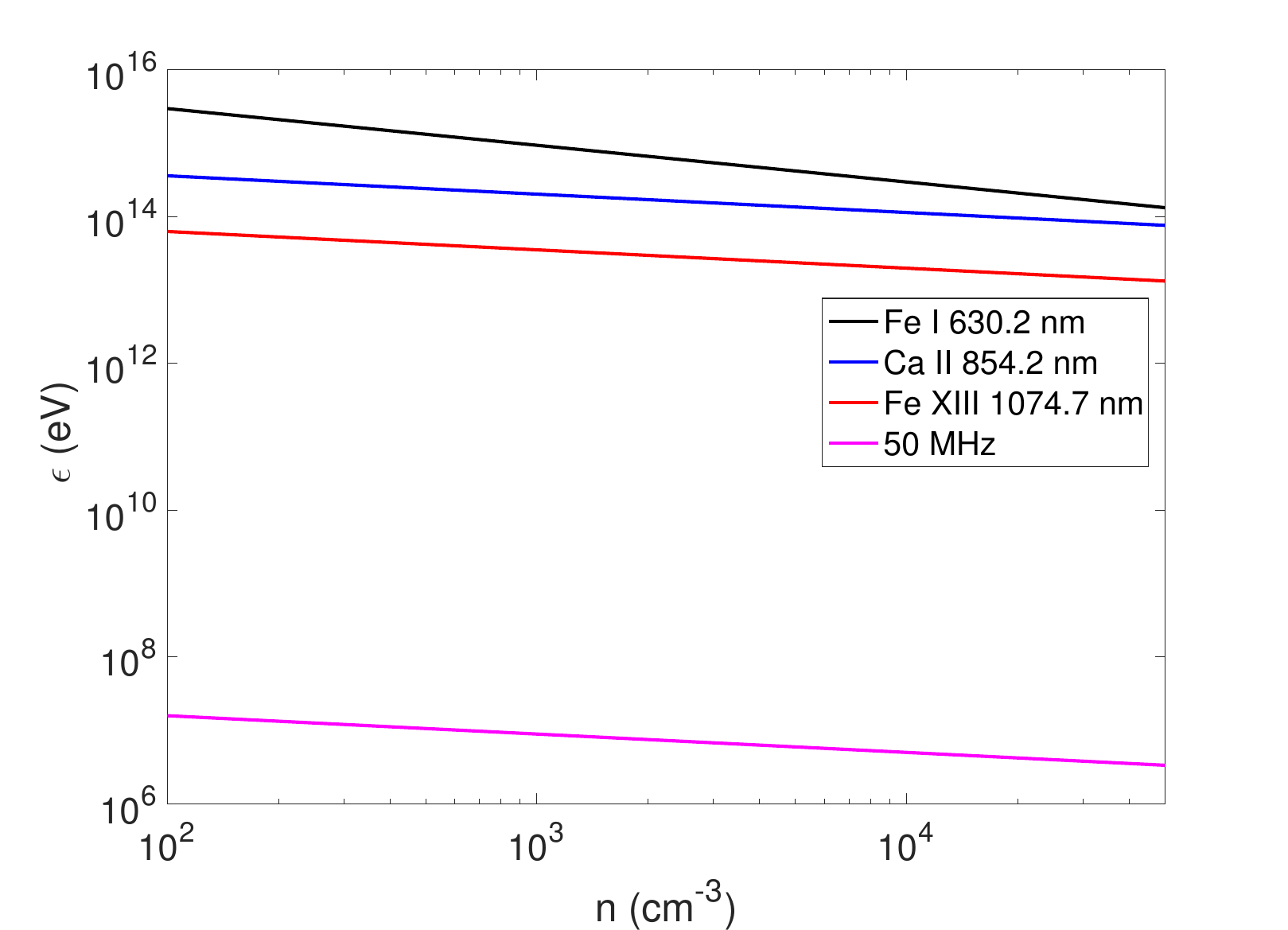}}
\caption{The plot of $\epsilon (n)$. The set of parameters is: $\lambda = 630.2$ nm, $B = 350$G, $Stokes\; V/I_c[\%] = 25$ (black); $\lambda = 854.2$ nm, $B = 350$G, $Stokes\; V/I_c[\%] = 1$ (blue); $\lambda = 1074.6$ nm, $B = 20$G, $Stokes\; V/I_c[\%] = 0.1$ (red); $f = 50$ MHz, $B = 350$G, $Stokes\; V/I_c[\%] = 50$ (magenta).}\label{fig1}
\label{fig2}
\end{figure}

\begin{figure}
\resizebox{\hsize}{!}{\includegraphics[angle=0]{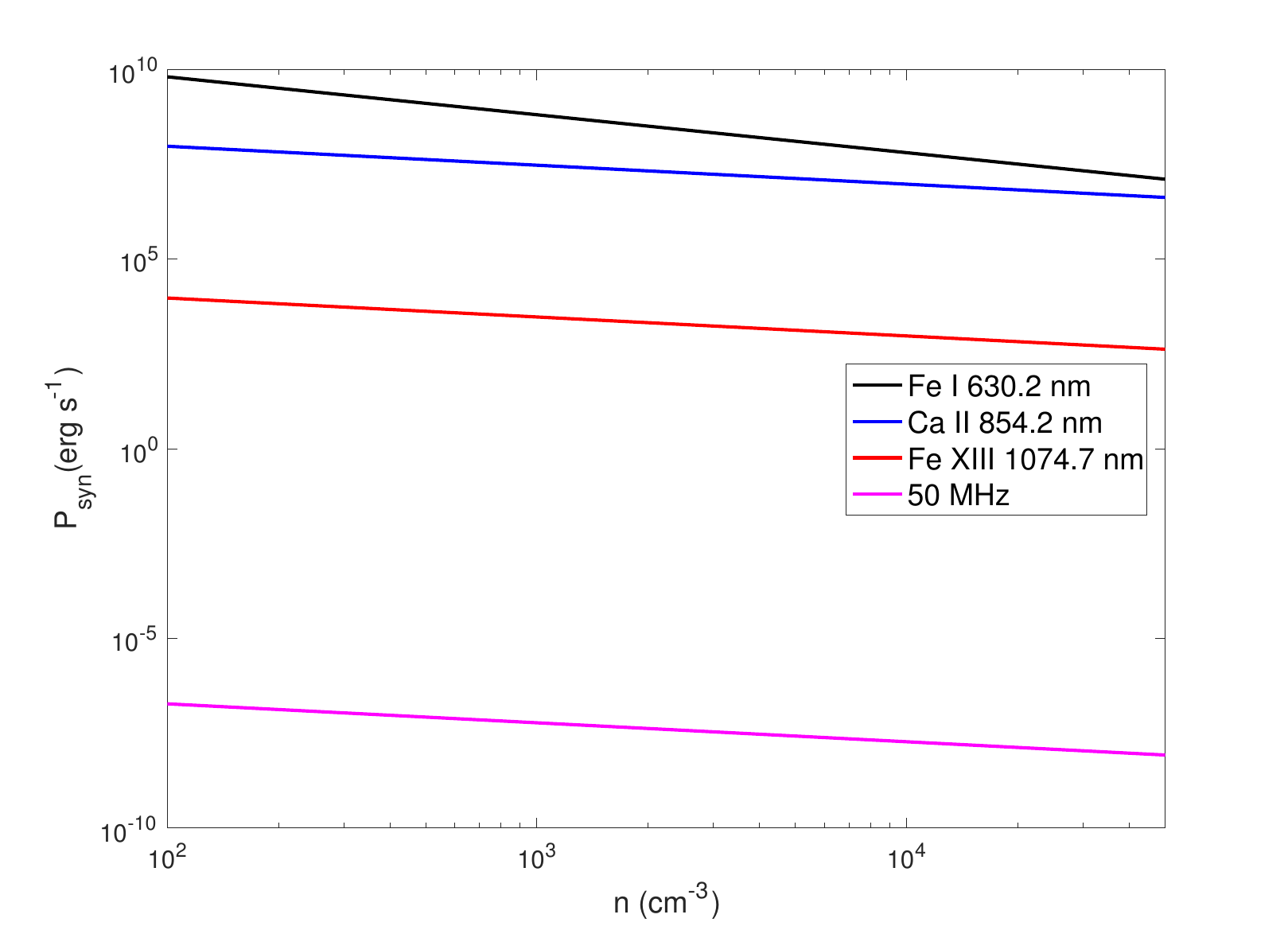}}
\caption{The plot of the synchrotron power versus the particle number density. The set of parameters is the same as in the previous figure.}\label{fig1}
\label{fig3}
\end{figure}

Highly relativistic  electrons might also cool in encounters with the ambient   gas particles; the resulting bremsstrahlung loss occurs on time   of $t_{br}\simeq4\times 10^7/n_p$ yr (\cite{ahar})with the proton number density (in the coronal region) $n_p \simeq10^9$ cm$^{-3}$. One can, show that the energy balance condition $t_{br}\simeq \gamma_{max}^{br}mc^2/\dot{E}$ leads to enormous Lorentz factors $\sim 10^{20}$ for Ca II 854.2 nm.  Similar situation pertains for other emission lines, therefore, the electron bremsstrahlung does not limit the process of electron acceleration.

From Eqs. (\ref{gama1},\ref{gsyn}), for the parameters given in the Table 1, one can show that the synchrotron mechanism is a limiting factor in all emission lines except $630.2$ nm, for which the maximum relativistic factor is given by Eq. (\ref{gama1}).

Plots of Maximum achievable energies $\epsilon = \gamma mc^2$, versus the number density are displayed in Fig2. These plots span the set of parameters: $\lambda = 630.2$ nm, $B = 350$G, $Stokes\; V/I_c[\%] = 25$ (black); $\lambda = 854.2$ nm, $B = 350$G, $Stokes\; V/I_c[\%] = 1$ (blue); $\lambda = 1074.6$ nm, $B = 20$G, $Stokes\; V/I_c[\%] = 0.1$ (red); $f = 50$ MHz, $B = 350$G, $Stokes\; V/I_c[\%] = 50$ (magenta). As evident from (see Eqs. (\ref{gama1},\ref{gsyn}), all plots reveal the maximum energy decreaseing  with $n$.  The line corresponding to $630.2$ nm has a different inclination because for it the maximum energy is not defined by the synchrotron losses. In Fig. 3, we show the synchrotron losses for the same set of parameters. We remind the reader that for Fe I $630.2$ nm, the maximum achievable energy is not limited by the synchrotron losses, and is defined by Eq. (\ref{gama1}); the synchrotron losses still take place.

The resonance acceleration of electrons with initial energies $\sim MeV$, therefore, might provide energies of the order $\sim 1.5\times 10^{7}$ eV for the radio band and $\sim 10^{14-15}$ eV for the rest, which means that the contribution of Sun in the generation of high and VHE energy cosmic rays must be taken seriously. However, one must proceed with caution and realize that total flux of solar cosmic rays may not be very large. First one must note that RWW operates on a very small subset of particles - the particles whose parallel (to the magnetic field) energy is much larger than the perpendicular component; the affected particles comprise a tiny sliver ( $\approx \omega_p/ \omega<<1$) of perpendicular energy distribution. Secondly even if all particles resonated, the flux of very high energy (VHE) particles in the vicinity of Earth $dN/(dtdA)\simeq FR_{\odot}^2/(r^2\epsilon$), will be very small. (here $R_{\odot}^2/r^2$ comes from the spherical symmetry and $r = 1 AU$ is the distance from the sun). 

But it is worth noting that the gyro-radius of the electrons with such high energies, $r_{gyro}\simeq\epsilon/(eB)\simeq 2.5\times 10^{10}$ cm, is smaller than $1 AU$ by three orders of magnitude, therefore, some of the particles might be significantly deflected from their original directions, which is a well known problem in cosmic ray physics. One finds that the time-scale of acceleration, $\tau=\epsilon/\dot{E}$ is by many orders of magnitude smaller than the escape time-scale, $\tau_{esc}\simeq R_{\odot}/c$. Thus the efficient acceleration of an individual particle as well as the continuous operation of the process is assured.  A detailed study of kinematics of particles is significant to estimate the flux, but it is beyond the scope of the present paper; we intend to study this particular problem in future.

Indirect signature of RWW could come from a secondary neutrino related process. High energy electrons, might produce neutrinos via the electron capture channel $e+p\rightarrow n+\nu_e$, which usually produces relatively low energy neutrinos, whereas the following process of $\beta$-decay $n\rightarrow p+e+\overline{\nu}_e$ produces high energy neutrinos. In particular, it is well known that the beta decay spectrum has a maximum intensity for the kinetic energy of anti-neutrino with the kinetic energy $\sim 1$ MeV \cite{beta} in the rest frame of neutron. On the other hand, neutron has almost the same energy as the proton (see the previous channel). If one considers the initial energy of electrons (before scattering) $\epsilon_0 = 10^{15}eV$, the Lorentz factor becomes of the order of $\epsilon_0/\epsilon_{n,r}\simeq 10^6$, where we have taken the neutron's rest energy, $\epsilon_{n,r}\simeq 1GeV$, into account. Therefore, the energy of anti-neutrino in the lab frame is $\gamma$-times bigger and equals $1$ TeV. Usually, from the sun one observes neutrinos in the TeV band, \cite{TeV}, where a contribution must come from the studied mechanism and not only from cosmic rays.

\section{Conclusions}

We have presented here our novel investigations of the solar corona; by studying the further energization of the already energetic coronal electrons (in a flare, for instance), we have explored a possible source of solar cosmic rays. The second stage of energization takes place by the resonant wave wave interaction (RWW) interaction between circularly polarized electromagnetic waves and the quantum waves of relativistic electrons. In particular, we selected the polarized radiation at $20-80$ MHz, Fe I 630.2 nm, Ca II 854.2 nm and Fe XIII 1074.7 nm that is used, often,  for coronal magnetic field diagnostics. 


Our results demonstrate that RWW interaction can catapult the relativistic electrons to much higher energies. Since the acceleration time-scales for electrons are much shorter than their escape times scales,  RWW interactions will be highly efficient, allowing particles to reach extremely high energies before they have a chance to escape from the corona.

By taking into account various (energy dependent ) processes , such as the synchrotron radiation, IC scattering and electron bremsstrahlung, that might limit RWW energization, we find that the polarized radiation 
can further energize the flare-generated relativistic electrons up to $10^{7}$ eV (for the radio band) and $\sim 10^{15}$ eV (for $20-80$ MHz, Fe I 630.2 nm, Ca II 854.2 nm and Fe XIII 1074.7 nm) (Figure~\ref{fig2})

Typical solar cosmic rays, known as SEPs, generally have lower energy than galactic cosmic rays. However, the presented study suggests that the Sun may produce VHE electrons approaching the energy of galactic cosmic rays.

The flux of such electrons could reach to $\sim 5\times 10^{-4}$ s$^{-1}$ cm$^{-2}$ during strong flare events
and might be detectable through remote measurements, including at ground level. Since the electrons might experience significant deflection while traveling through the coronal magnetic field, the detection could be  very challenging.

In a follow-up study, we plan to investigate the RWW interactions for protons, another primary component of solar cosmic rays.

\section*{Acknowledgments}
The research was supported by a German DAAD scholarship within the
program Research Stays for University Academics and Scientists, 2024 (ID: 57693448), and also was  supported by Shota Rustaveli National Science Foundation of Georgia (SRNSFG) Grant: FR-23-18821. ZO is grateful to prof. G. Dvali for fruitful discussions and comments. ZO acknowledges Max Planck Institute for Physics (Munich) for hospitality during the completion of this project.

\nolinenumbers





\end{document}